\documentclass[]{spie}  
\usepackage[]{graphicx}

\title{The system software development for prime focus spectrograph on Subaru Telescope}

\author{
Atsushi Shimono\supit{a},
Naoyuki Tamura\supit{a},
Hajime Sugai\supit{a},
Hiroshi Karoji\supit{a}
\skiplinehalf
\supit{a} 
  Kavli Institute for the Physics and Mathematics of the Universe, The University of Tokyo, Japan; \\
}

\authorinfo{Further author information: (Send correspondence to A.S.)\\
A.S.: E-mail: atsushi.shimono@ipmu.jp, Telephone: +81 4 7136 5976\\
}

\begin{document} 
  \maketitle 

\begin{abstract}

The Prime Focus Spectrograph (PFS) is a wide field multi-fiber
spectrograph using the prime focus of the Subaru telescope, which is
capable of observing up to 2400 astronomical objects simultaneously.

The instrument control software will manage the observation procedure
communicating with subsystems such as the fiber positioner "COBRA",
the metrology camera system, and the spectrograph and camera systems.
Before an exposure starts, the instrument control system needs to
access to a database where target lists provided by observers are
stored in advance, and accurately position fibers onto astronomical
targets as requested therein.  This fiber positioning will be carried
out interacting with the metrology system which measures the fiber
positions. In parallel, the control system can issue a command to
point the telescope to the target position and to rotate the instrument
rotator. Finally the telescope pointing and the rotator angle will be
checked by imaging bright stars and checking their positions on the
auto-guide and acquisition cameras.
After the exposure finishes, the data are collected from the
detector systems and are finalized as FITS files to
archive with necessary information.

The observation preparation software is required, given target lists
and a sequence of observation, to find optimal fiber allocations
with maximizing the number of guide stars.
To carry out these operations efficiently, the control system will be
integrated seamlessly with a database system which will store
information necessary for observation execution such as fiber
configurations.

In this article, the conceptual system design of the observation
preparation software and the instrument control software will be
presented.

\end{abstract}



\section{INTRODUCTION} \label{sec:intro}

The Prime Focus Spectrograph (PFS \cite{2012arXiv1206.0737E}) 
is a wide field multi-fiber
spectrograph using the prime focus of the Subaru telescope, which is
capable of observing up to 2400 astronomical objects simultaneously.
Also the PFS is assumed to be used for large survey programs up to 
300 nights in 5 years in total. 
The PFS system software is required 
to coordinate science observations and control entire instrument 
with interfacing control software of each component 
under environment conditions of Subaru telescope. 
For observations of survey programs, 
a number of targets observed at once is significantly large, 
capabilities for on-site automatical rearrangements and redesignings of 
survey programs are important for efficient survey operation. 
Since the PFS is planned to be an open use instrument of the Subaru telescope, 
the PFS system software and its architectural design is required to support 
both of survey programs and classical mode observation programs.

%

In this article, we will present 
requirements, system operational constraints, 
and conceptual design of the PFS system software.

\section{Overview of PFS System Software}

The PFS System software divides into three main deliverables (packages): 
the Observation Preparation Software, 
the Observation Execution Control Software, 
and the Engineering software.

The Observation Preparation Software will be 
used for the preparation of observations by astronomers at their local. 
This software package is required to perform: 
to divide large target lists into observations, 
generate the fiber and COBRA assignment as data files with 
matching targets to fibers with suitable guide star assignments, 
and 
generate observation sequence scripts for Subaru Observation System (Gen2) 
in a suitable format.

The Observation Execution Control Software will 
run at the Subaru observatory to 
control, configure and manage the entire PFS subsystem control software 
by communicating with and receiving commands from Gen2. 
For survey observation program, to achieve better survey performance, 
this software package will be required to work in cooperation with 
the data reduction pipeline and the observation preparation software. 

The Engineering software is used for test, maintenance and calibration.
This package 
provides engineering software per each subsystem which can work only 
with the local system, 
and 
simulator for each connection and interaction point 
for simulating missing subsystems during development and maintenance.

\subsection{Relation among system software packages}

To make three packages easily in cooperation, 
we will have one Relational Data Base Management System (RDBMS) database 
shared among packages. Every instrument configuration parameters and 
observation sequence information will be stored and shared in this database. 

The engineering software package will be used independently with the Subaru 
telescope control system (TCS) and be used for local test, maintenance, and 
calibration. These results and calibrated parameters will be stored into 
the shared database and be used from other software packages. 
The observation preparation software will be used by astronomers to make 
an operation sequence for each observation program 
which will be supplied to the observation execution control software 
package via the shared database.

The observation execution control software package will control entire PFS 
sub-systems to perform science or engineering observations with referring 
information in the shared database supplied by other packages. 
For survey programs, the observation execution control software will 
communicate with the on-site quick-look data reduction system and 
the observation preparation software just after each exposure 
to update observation sequences of survey program considering quality 
assurance on observed data.

\subsection{Relation to the Subaru}

Among the entire PFS system software packages, mainly the observation 
execution control software will be required to communicate and cooperate 
with the Subaru telescope control software.

\begin{figure}
\begin{center}
\includegraphics[scale=0.9]{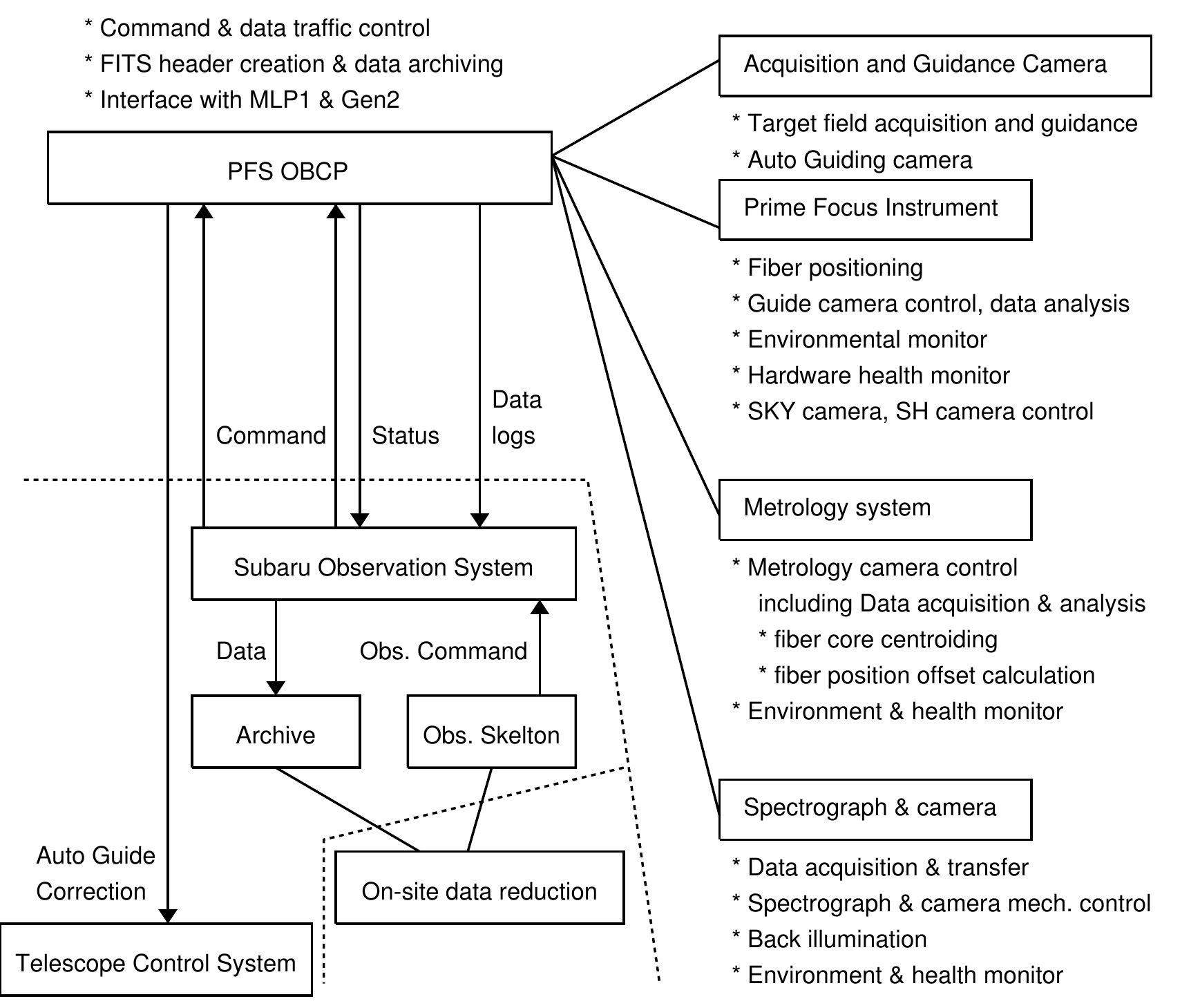}
\end{center}
\caption
{\label{fig:overview}
Overview of relationship from the Observation Execution Control Software of 
the PFS system software are shown. 
}
\end{figure}

This package must run within environments of the Subaru observatory 
under three constraints, 
1) this package must well communicate with the Subaru TCS and Gen2,
2) this package must send the Acquisition and Guide correction 
information to the Subaru TCS via a communication link specified by the 
Subaru observatory, 
3) this package must perform operations under the operational conditions by 
the software environment of the Subaru telescope and the Gen2, 
as summarized in Figure.~\ref{fig:overview}. 

The observation execution software package (or the observation preparation 
software package) will create and supply observation sequence files 
named as skeleton or operational file. 
These sequence files will be executed by the Gen2 during the observation run 
with interfacing the other telescope control software and the PFS observation 
execution control software package (generally called as OBCP in the Subaru 
telescope) using three communication channels: 
1) A command channel from Gen2 to OBCP, 
used by Gen2 to execute commands or query status. 
Every single command of OBCP and subsystems should be executable from Gen2. 
2) A command channel from OBCP to Gen2 for querying OCS status, 
used by OBCP to query status from Gen2 with their name. 
3) A data channel from OBCP to Gen2, 
used by OBCP to transfer observed data from each exposure 
in the format required by the Subaru observatory.
Gen2 will send the received data into 
the Subaru Telescope data ARchive System (STARS).

\subsection{Relation to subsystem control software}

Every PFS subsystems supply their control software, 
the observation execution control software package will interface with 
these sub-system control software by three communication channels: 
1) A command channel from the observation execution control software 
to subsystems to configure subsystems and execute commands 
2) A status and log channel from subsystems to the observation execution 
control software to archive system statuses and logs of subsystems 
3) A data channel from subsystems to the observation execution control software 
to transfer observed data by detectors' control software. 
Examples of these communication channels are shown in 
Figure.~\ref{fig:overview}.

\section{System Architectural design}

\subsection{Activities taken by software packages}

\begin{figure}
\begin{center}
\includegraphics[scale=0.8]{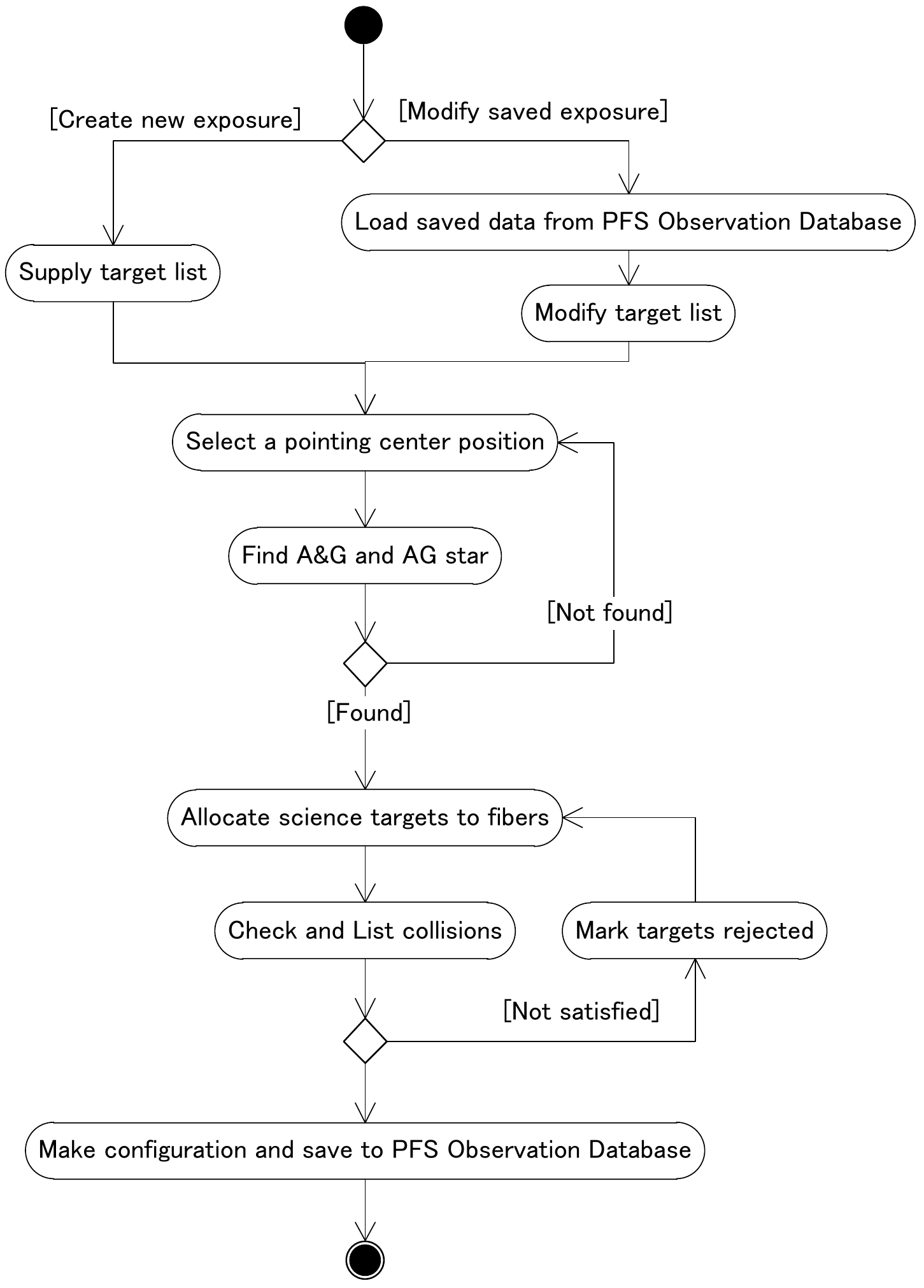}
\end{center}
\caption
{\label{fig:obs-prep}
Activity diagram for preparing single observation.
}
\end{figure}

\begin{figure}
\begin{center}
\includegraphics[scale=0.8]{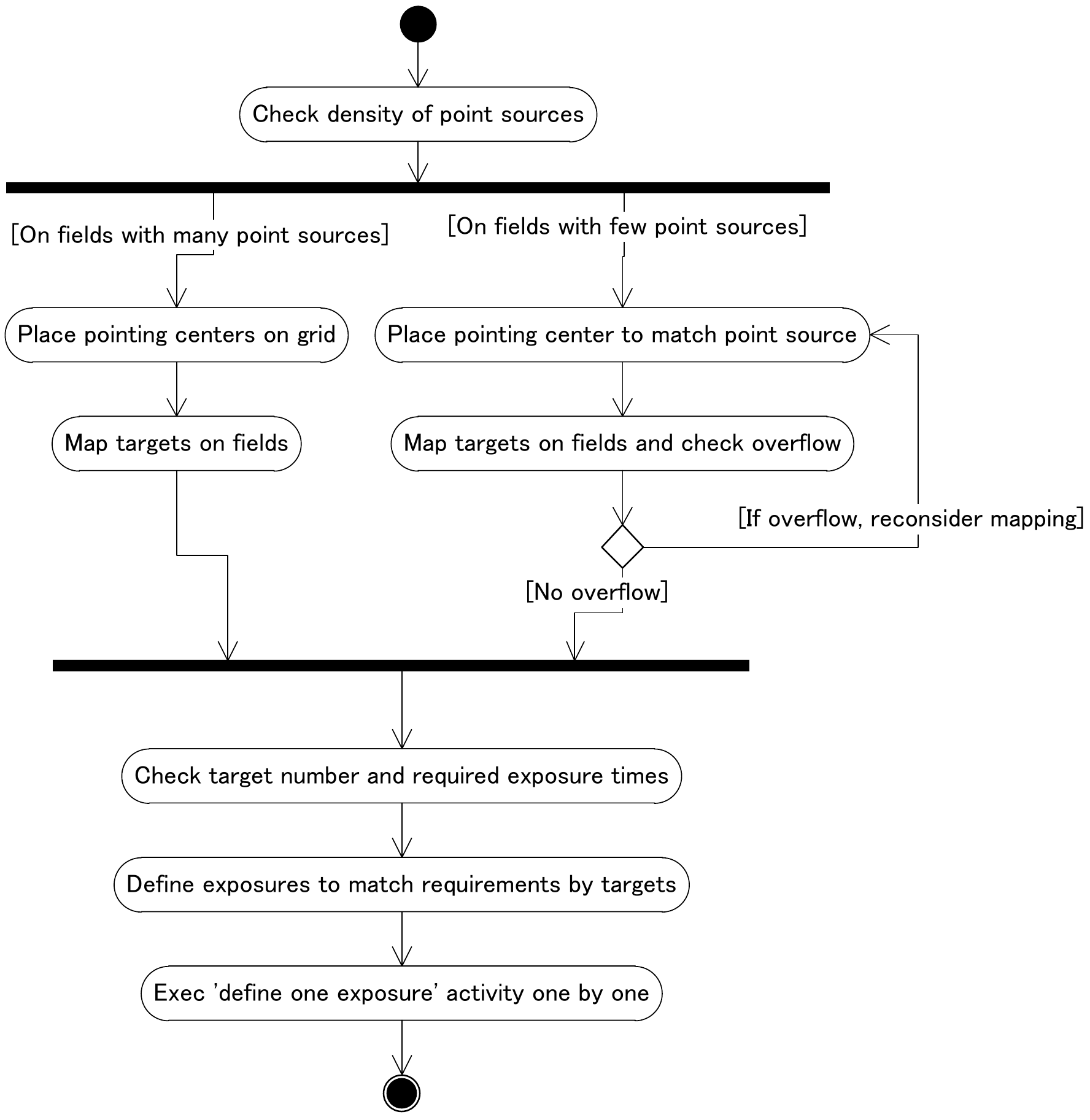}
\end{center}
\caption
{\label{fig:survey-prep}
Activity diagram for preparing observation programs of entire survey program.
}
\end{figure}

Activities performed by the observation preparation software are mainly 
to prepare a single observation (Figure.~\ref{fig:obs-prep}) and 
to prepare observation programs for entire survey programs 
(Figure.~\ref{fig:survey-prep}). 
The former activity will perform to define and update any required information 
for a single observation by the observation execution control software 
from a supplied list of targets by an astronomer. 
Since possible areas and brightness of objects used for Acquisition and 
Guidance (A\&G) are limited in corresponding to a pointing center position 
of the instrument, 
the software need to check existence of suitable objects for A\&G 
consulting with guide star catalogs. 
When the software could not find suitable objects more than required number, 
a user will be warned and be noticed to find another possible pointing 
center position by using the software. 
Once the pointing center position is confirmed, the software will try 
to allocate supplied science and calibration targets to fibers, considering 
distortion by the wide field collector (WFC) and collision among positioners. 
The latter activity uses the former one, and will be used to 
divide supplied target lists for survey programs into series of single 
observations 
considering target spatial density, required signal to noise ratio per target, 
and so on. 
This activity will also be used at the observatory during survey observations 
to update observation sequences by considering results from quick-look 
data reduction. 

\begin{figure}
\begin{center}
\includegraphics[scale=0.9]{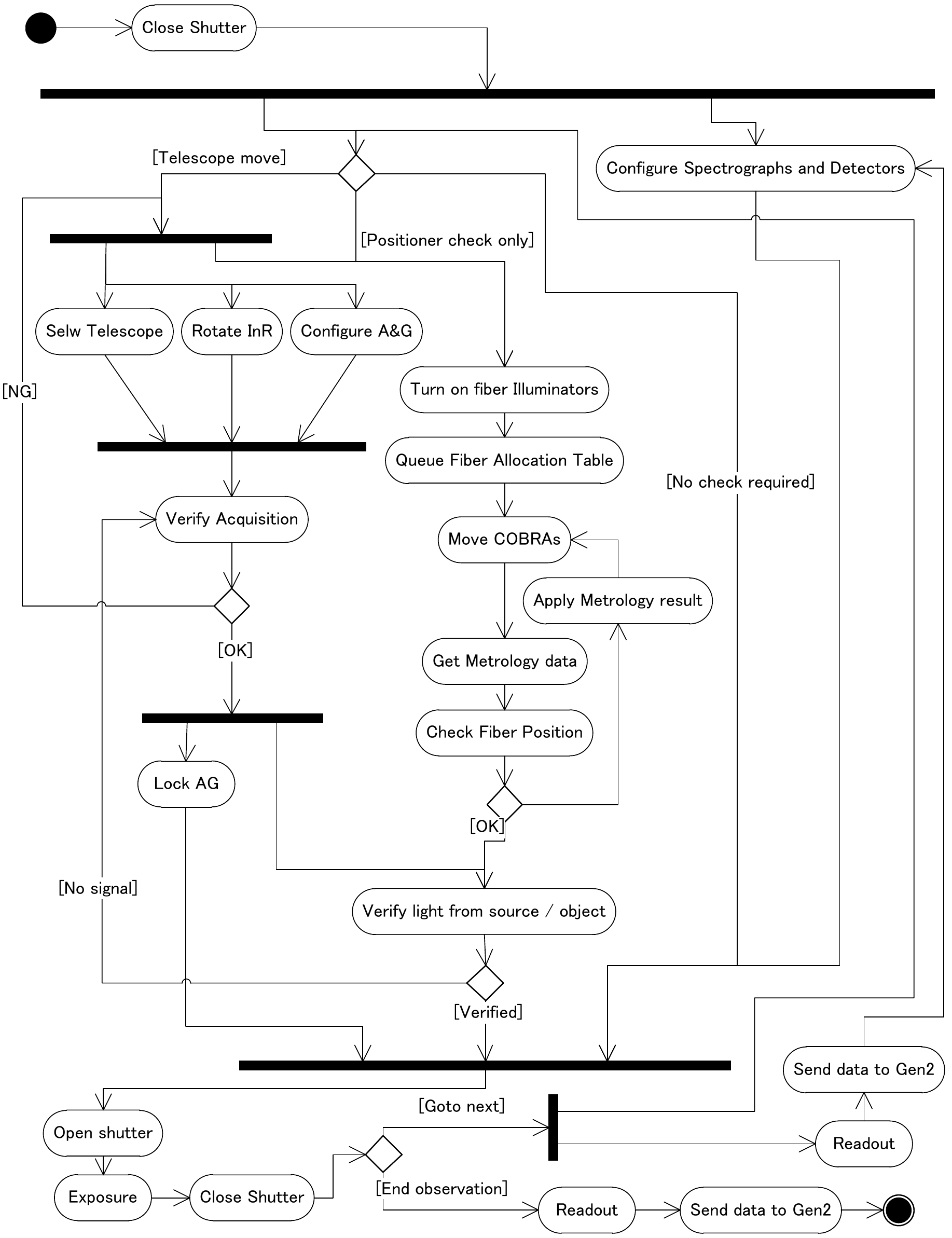}
\end{center}
\caption
{\label{fig:obs}
Activity diagram for executing a series of observations.
}
\end{figure}

The observation execution control software will perform every activity required 
for observations as shown in Figure.~\ref{fig:obs} and also every user 
interface, 
including: 
providing integrated graphical user interfaces to command and check 
the fiber systems including the back illuminator; 
sequencing the Metrology process until science and calibration targets 
have been successfully acquired, and target exposures; 
completing the data acquisition by assembling required FITS header, 
and interfaces with the data handling system of Gen2; 
reporting errors and arises alarms of the entire instrument including 
the subsystems in the required way; 
creating observation and system logs including system statuses 
received from subsystems; 
and
providing logging systems which enable operators and engineers 
to access, handle, and search easily even while offline.

To make survey observations more effectively and overhead more smaller, 
activities for target acquisition including telescope pointing and metrology 
procedure on positioners, and 
activities for finalizing data into the FITS format and configuration of 
spectrographs will be performed in parallel.

\subsection{Shared database}

To share configuration parameters for the instrument itself and observations, 
we will share one RDBMS database among every software components. 
Also we will use the Extensible Markup Language (XML) format to store 
and exchange its data for the observation preparation software working at 
astronomers' site. 

For survey programs, updating observation parameters including list of targets 
for upcoming exposures and their order depending observation conditions 
and observed data qualities is planned. 
To perform these on-site, we need to share information of targets and their 
observed data quality between the observation preparation software and 
the on-site quick-look data reduction pipeline. 
For this purpose, we will have another on-site database to store list of 
survey targets with required and observed data quality per each survey 
programs, including a copy of parameters used for executed observation.

\subsection{Messaging hub, status and message log service}

For communication between the PFS system software and every subsystem control 
software, we will select and use one network messaging library. 
Also, we are planning to have a log service both for message over network 
and for instrument status smoothly integrated to the network messaging 
library.

\subsection{Error detection and handling}

To have one central log service both for message exchanged over network and for 
instrument status with secure timestamps 
enables us to view and check those integrated log over entire instrument
both on-line and off-line, 
and also to have another service independent from instrument control system 
to alert operators when any instrument status goes wrong or a continuous error 
detected for messages.

%
%
%
%
%

\section{DISCUSSION}

For on-site data storage used for survey observations, 
we considered two way; 
1) totally integrated database system for all required information including 
observation configuration, target objects, and their achieved data quality from 
on-site data reduction pipeline, 
2) separated database system for observation configuration and 
for list of target objects and their achieved data quality by each survey 
program. 
Since the PFS is supposed to be delivered to the Subaru telescope as a future 
facility instrument, we need to consider both of our proposing survey programs 
and also PI typed observation programs including small survey programs called 
as an intensive observation. 
Taking former option makes system software package more complex 
when we consider to ensure secure separation among observation programs, 
and also makes operational costs and risks on switching modes larger. 
Therefore we would take having a somewhat independent software suite including 
database system for survey programs.

\acknowledgments     

We gratefully acknowledge support from the Funding Program for World-Leading 
Innovative R\&D on Science and Technology (FIRST) "Subaru Measurements of 
Images and Redshifts (SuMIRe)", CSTP, Japan.


\bibliography{spie-pfs-syssoft}   
\bibliographystyle{spiebib}   

\end{document}